\documentclass[%
 reprint,
 amsmath,amssymb,
 aps,
]{revtex4-2}

\usepackage{graphicx}
\usepackage{dcolumn}
\usepackage{bm}
\usepackage{amsthm}

\begin{document}


\title{Quantum Finite Volume Method for Computational Fluid Dynamics \\with Classical Input and Output}

\author{Zhao-Yun Chen}
\author{Cheng Xue}%
\author{Si-Ming Chen}
\author{Bing-Han Lu}
\author{Yu-Chun Wu}
 \email{wuyuchun@ustc.edu.cn}
\affiliation{%
 Key Laboratory of Quantum Information, CAS
}%
\affiliation{University of Science and Technology of China}%


\author{Ju-Chun Ding}
\author{Sheng-Hong Huang}
 \email{hshnpu@ustc.edu.cn}
\affiliation{
 Department of Modern Mechanics, USTC
}%

\author{Guo-Ping Guo}
 \affiliation{Key Laboratory of Quantum Information, CAS}
 \affiliation{University of Science and Technology of China}%
 \affiliation{
 Origin Quantum Computing, Hefei
}%
 \email{gpguo@ustc.edu.cn}


\date{\today}

\begin{abstract}
Computational fluid dynamics (CFD) is a branch of fluid mechanics that uses numerical methods to solve fluid flows. The finite volume method (FVM) is an important one. In FVM, space is discretized to many grid cells. When the number of grid cells grows, massive computing resources are needed correspondingly. Recently, quantum computing has been proven to outperform a classical computer on specific computational tasks. However, the quantum CFD (QCFD) solver remains a challenge because the conversion between the classical and quantum data would become the bottleneck for the time complexity. Here we propose a QCFD solver with exponential speedup over classical counterparts and focus on how a quantum computer handles classical input and output. By utilizing quantum random access memory, the algorithm realizes sublinear time at every iteration step. The QCFD solver could allow new frontiers in the CFD area by allowing a finer mesh and faster calculation.
\end{abstract}

\maketitle

\section{Introduction}
Computational fluid dynamics (CFD) is the area that utilizes numerical methods to obtain the physical properties of fluids. It has many applications, such as aid in designing aircraft or automobile. CFD is often related to solving a series of partial differential equations (PDEs) and can compute the evolution of physical characteristics of fluid at a given space, including density, momentum, and energy. These characteristics would provide us essential references for the properties of the fluid in the computational space. There are three typical physical governing equations of the CFD: Navier--Stokes (NS), Euler, and Reynold--Averaged Navier--Stokes (RANS) equations.

The finite volume method (FVM) is a typical numerical method to discretize these physical equations. In FVM, the computational space is discretized into small cells by dense grid points, separately solving every cell's evolution at a small timestep and finally integrating all time steps. The PDE will be converted to a sparse linear equation at every time step, whose dimension $N$ has a linear dependency on the number of cells. In practice, sparse matrix linear solvers such as the conjugate gradient method are available. The best time complexity of the conjugate gradient method is $O(Ns\kappa \log 1/\epsilon)$ time complexity where $s$ is the sparsity number (the maximum number non-zero elements in each row or column), $\kappa$ is the condition number, and $\epsilon$ is the precision \cite{Shewchuk1994An}. A typical problem for the FVM is that when the problem size grows large, the computing resources will become expensive. 

Instead of using classical computers, quantum computing is a promising computing paradigm that offers exponential acceleration over classical computing approaches. Many quantum algorithms, including quantum factorization\cite{shor1999polynomial}, quantum simulation\cite{georgescu2014quantum, Berry2014Proceedings, berry2015hamiltonian, o2016scalable}, and the linear system solvers, \cite{PhysRevLett.103.150502, ambainis2010variable, Childs2017Quantum} have already appeared to prove this idea. Thus, we try to accelerate a CFD solver with quantum computing. There were some works about solving linear PDEs using a quantum computer\cite{2002.07868, 1705.09361, 1711.05394, 1207.2485}. However, these methods cannot be directly applied to solve CFD because the Navier--Stokes equation is a non-linear PDE, which is not covered by these previous results.

This paper introduces a quantum solver for CFD problems (QCFD solver) based on classical FVM. We show that with only classical input, the time cost of each iteration step can be reduced to polylogarithmic dependency on the problem size. This provides an exponential speedup of the FVM. The QCFD solver can fully reproduce the result of the FVM. With the output given classically at every time step, the QFVM is capable of the steady or unsteady problem with similar configurations.

To apply quantum algorithms to practical problems, the conversion between the classical and the quantum data could become a bottleneck, especially when using the quantum linear solver (QLS) as a submodule\cite{AaronsonRead}. In our algorithm, the input and output are all classical data. To achieve this, we design a quantum memory layout based on quantum random access memory (QRAM) \cite{giovannetti2008architectures, giovannetti2008quantum}. At the input stage, the memory layout helps to implement subprocedures required by the QLS. At the output stage, we sample the output state and update the memory classically. We show that these two processes, which act as the interface between classical and quantum data, can both run in polylogarithmic time. They enable us to integrate the quantum linear solver submodule into the classical FVM to achieve speedup. The time complexity of our algorithm is calculated by scaling the time between two iteration step, which is
\begin{equation}
    O\left(
    \frac{(s^3+\log N)s\kappa\log^3 N}{\epsilon^2}\mathrm{polylog}(s\kappa/\epsilon)
    \right).
\end{equation}

Our algorithm can be compared to the classical algorithm directly. The algorithm has classical input and output and uses the same definition of condition number and the error threshold. For the condition number problem, we implement a quantum version of the Jacobi preconditioner and integrate it into the memory layout. It has the same effect as applying a Jacobi preconditioner, which is common in the classical FVM. Therefore,  both quantum and the classical algorithm have a linear dependency (if we ignore the polylogarithmic term) on the same condition number. We also provide evidence that the quantum and the classical error threshold are the same under the sense of $l_\infty$ norm. 

As a result, compared to the time complexity of the classical counterpart, the quantum solver runs faster in terms of the problem size $N$, but slower in the dependency on the precision with a quadratic term. Our algorithm will have better performance when the problem size is large enough, i.e., $N\gg 1/\epsilon^2$. We performed a numerical simulation on the Onera M6 test case showing that the algorithm can output correctly with such a condition.

\section{Preliminaries}

\subsection{Discretization and Linearization of the Physical Governing Equation}\label{LinearizationNS}

The typical physical governing equations (Euler, NS, RANS) have to be linearized to apply to the FVM. In this paper, we do not focus on the detail of the linearization. Instead, we apply the identical linearization method to the classical algorithm and analyze the relationships among the equation variables.

Here we take a two dimensional NS equation with compressible flow as an example. First write down the differential form of the NS equation:
\begin{equation}
    \frac{\partial}{\partial t}\int_\Omega UdV+\oint_{\partial\Omega}\bm{F} \cdot d\bm{S}=0,
\end{equation}
where 
\begin{equation}
\begin{array}{ccc}
U = \left[
\begin{array}{c}
\rho\\
\rho u\\
\rho v\\
\rho E
\end{array}
\right]
&
\bm{F}_x = \left[
\begin{array}{c}
\rho u\\
\rho u^2 + p\\
\rho u v\\
\rho u H
\end{array}
\right]
&
\bm{F}_y = \left[
\begin{array}{c}
\rho v\\
\rho u v\\
\rho v^2 + p\\
\rho v H
\end{array}
\right],
\end{array}
\end{equation}
for any volumne $\Omega$ and its boundary $\partial\Omega$.

\begin{figure}
    \centering
    \includegraphics[width=\linewidth]{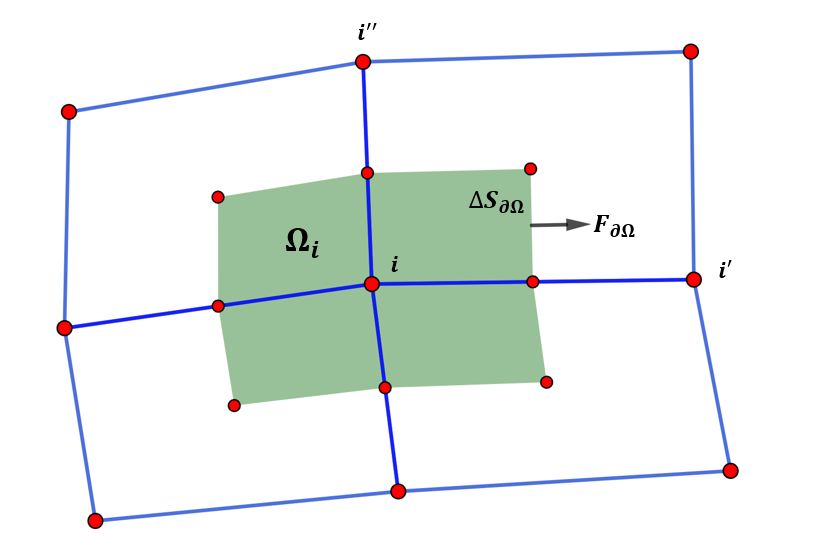}
    \caption{The grid cell around $i^{\rm th}$ point. $\bm{F}_{\partial \Omega}$ is the flux at the certain boundary; $\Delta S_{\partial\Omega}$ is the area. $\Omega_i$ is the volume of this cell.}
    \label{fig:gridcell}
\end{figure}

To discretize it spatially and timely, we split the space and time into small grid cells. At the cell i and time step n, the NS equation can be discretized to
\begin{equation}\label{EqnDiscretized}
    \frac{\Omega_i}{\Delta t}\left(U^{n+1}_i-U^n_i\right)=-\sum_{\partial\Omega}\bm{F}_{i,\partial\Omega}^{n+1}\cdot\Delta\bm{S}_{i,\partial\Omega}^{n+1},
\end{equation}
where implicit Euler method is applied. We define the right hand side of the equation~(\ref{EqnDiscretized}) as the residual of this point, denoted by $R_i^{n+1}$. The $\bm{F}_{i}$ is defined by the difference scheme, which is calculated by variables $\vec{U}$ in the surrounding cells. The difference scheme gives a relation between nodes. In this paper we define a matrix $C$ which has
\begin{equation}
    C_{i,i'} = 1 
\end{equation}
if i and $i'$ are related in the difference scheme. In other words, $C_{i,i'}=1$ means calculateing the residual at $i^{\rm th}$ node uses the variables in $j^{\rm th}$ node. Specially, we always have $C_{i,i}=1$. 

Let $\Delta \vec{U}^{n+1}=\vec{U}^{n+1}-\vec{U}^n$, we have
\begin{equation}
    \left(\frac{\Omega_i}{\Delta t}\delta_{i,i'}+\left.\frac{\partial R_{i,k}}{\partial U_{i',k'}}\right|_{U=U^n}\right)\Delta U^n_{i',k'}=-R^n_{i,k}.
\end{equation}

Simply replacing $\bm{A}=\left(\frac{\Omega_i}{\Delta t}\delta_{i,i'}+\left.\frac{\partial R_{i,k}}{\partial U_{i',k'}}\right|_{U=U^n}\right)$, we obtain a linear equation whose solution implies the time evolution of the physical variable $\vec{U}$.

The coefficient matrix $\bm{A}$ is a sparse matrix. From equation~(\ref{EqnDiscretized}), the $A_{i,k,i',k'}$ is non-zero when $i$ and $i'$ are related in the difference scheme ($C_{i,i'}=1)$. The sparse number (number of nonzero element in a row or column) is fixed by how we select the difference scheme, denoted by $s$.

Regardless of the physical governing equation, the discretization and the linearization following the classical FVM method do not change. We will finally show that the spatial or time difference scheme does not affect how this algorithm works, and only the constant-coefficient will change in the analysis of the time complexity.

\subsection{Quantum Algorithm with Classical Input and Output}

When we use a quantum computer to cope with a practical problem, we should always expect that the input and output are classical. Many quantum algorithms have been proposed and claimed to be faster (exponentially or polynomially) than their classical counterparts. However, a large portion of them only beats classical algorithm under some theoretical limitations. A typical example is the famous quantum linear system algorithm: Harrow-Haddism-Lloyd (HHL) algorithm\cite{PhysRevLett.103.150502}, which can prepare the state $|x\rangle$ encoding the solution of the linear equation $Ax=b$. This algorithm uses $O(\log N)$ calls to linear equation oracles, where the classical counterpart has to perform at least $O(N)$ calls. Based on the ability to accelerate solving a linear equation, many quantum machine learning algorithms were proposed and accelerated exponentially over the classical algorithm. However, most of these algorithms did not answer how to deal with real-world data or obtain a classical output. In this paper\cite{AaronsonRead}, Aaronson raised a series of obstacles for applying the HHL algorithm on quantum machine learning algorithms with real-world data. The main problems include how to input the classical data into the quantum computer and extract information from the output state given by the HHL algorithm. If we hope to preserve quantum speedup, two operations are forbidden. One is to prepare the input state $|b\rangle$ with an encoded quantum circuit, where even reading all data entries requires $O(N)$ time; the other is to perform sampling on the output state to extract the state to a classical vector with $O(N)$ times measurement. 

We believe the obstacles that appeared in ``HHL-based'' quantum machine learning algorithms are even more challenging if we want to accelerate the FVM for CFD problems quantumly. In the classical method, the physical governing equation (e.g., Navier--Stokes equation) will be discretized in both time and space. At every time step, we linearize the physical governing equation to a linear equation, and its solution represents the evolution of the physical variables. Finally, we can integrate all time steps to obtain a stable solution. The time complexity of the FVM mainly depends on the time complexity of the linear solver, which is at least $O(N)$. It is straightforward to suppose that if we change the linear solver to a quantum version, we will have a quantum accelerated CFD solver. However, the two problems mentioned above exist in every step of the time integration. One is how to generate the input from the physical variables at one time step; the other is how to update the physical variables from the quantum linear solver's output. In conclusion, the quantum algorithm will fail to demonstrate quantum advantage if time complexity requires an extra $O(N)$ multiplier.

Our proposed QFVM algorithm will consider these obstacles. We assume the input and output of this algorithm are all classical, ensuring that this algorithm can be run in the quantum computer without providing more input than the classical algorithm.

\subsection{Quantum Random Access Memory}\label{PreliQRAM}
Quantum random access memory is the storage device for the quantum computer. As the quantum analog of RAM, QRAM allows a quantum computer to obtain classical data with given addresses in quantum parallel. In other words, QRAM could perform such unitary transformation:
\begin{equation}
    U_{\rm QRAM}|i\rangle_A|0\rangle_D = |i\rangle_A|d_i\rangle_D,
\end{equation}
where A and D denote the address and the data registers. $d_i$ is a classical data entry stored at the address $i$.

A seminal architecture called ``bucket-brigade'' provides an efficient way for querying. There have been many proposed physical implementations of such architecture, such as optical system \cite{giovannetti2008architectures}, acoustics system \cite{hann2019hardware}, and circuit quantum electrodynamics\cite{naik2017random}. Our work is based on the QRAM with architecture implemented by any of the physical systems. To eliminate the difference in understanding the availability of the QRAM, we list all assumptions when we apply the QRAM to our algorithm.

First, the QRAM is general to all input addresses and their superpositions, namely $\sum c_i|i\rangle$. The QRAM should be an arbitrary data loader rather than only allowing to prepare the $\sum|i\rangle|d_i\rangle$ state.

Second, if the address register has been prepared, performing one query costs $O(\log N)$ time where the full data length is N.

Third, we assume a QRAM has at least a classical RAM capability, enabling access to a single entry or overlaying it to another value with constant time. Meanwhile, the QRAM should be compatible with a classical computer. A classical computer can read the data in QRAM without extra cost.

Even though a real physical implementation of QRAM is hard, these assumptions are reasonable because they do not exceed the capabilities of the previous physical implementations. 

We claim our algorithm as ``classically input and output'' under the sense that the input and output of the algorithm are stored in the QRAM. Because we believe in the compatibility of QRAM and classical computer, the problem definition, data initialization, and post-processing of the calculation results can all be performed in a connected classical computer.

\section{Quantum Finite Volume Method}\label{QFVMsec}

This section introduces the quantum finite volume method (QFVM) for CFD problems based on the implicit Euler method and classical FVM. As described in the preliminary section (\ref{LinearizationNS}), the Euler, NS, or RANS equation can be discretized timely and spatially and finally linearized following specific classical methods. We do not focus on these methods but only transplant them into a quantum version based on the theory that any classical function can be implemented in a quantum computer.

Take a two-dimensional Euler equation with compressible gas (density, X- or Y-directional momentum and energy) as the example. The physical variable $\vec{U}$ contains the physical properties of all grid cells. We discretize the computing space into N grid cells, then the size of $\vec{U}$ is 4N. We use a two-level subscript $(i,k)$ to identify a single element in $\vec{U}$ by $U_{i,k}$, where $i$ denotes the $i^{\rm th}$ grid cell and $k$ denotes the $k^{\rm th}$ physical characteristics.

The time evolution of $\vec{U}$ is realized by solving the linearized NS equation using implicit Euler time iteration scheme. The calculation is iterative, and we use a subscript n to denote the $n^{\rm th}$ iteration step. At any time step, the equation has a general form:
\begin{equation}
    \bm{A}^n(\vec{U}^n)\Delta\vec{U}^{n+1}=-\vec{R}^n(\vec{U}^n),
\end{equation}
where $\bm{A}$ is the Jacobian matrix and $\vec{R}$ is the residual vector.  These two parts are determined by the physical variable $\vec{U}$ at $n^{\rm th}$ step. The unknown is $\Delta\vec{ U^{n+1}}$. The system evolves one step by calculating
\begin{equation}
    \vec{U}^{n+1} = \vec{U}^n + \Delta\vec{U}^{n+1}.
\end{equation}

Similar to the definition of $U_{i,k}$, the matrix element of $\bm{A}$ is denoted by $A_{i,k}^{i',k'}$. Element in $\vec{R}$ is denoted by $R_{i,k}$. 

The initial state of the physical variable $\vec{U}^0$ should have been stored in QRAM. It is given as the input of the algorithm. Another part of the input is about the spatial grid points, which split the space into cells. These data should also be stored in the QRAM, ready for quantum query.

Our work mainly concentrates on how to bridge the gap between quantum processes and classical processes. On top of our work, we design a QRAM-based memory layout inspired by previous works\cite{Kerenedis2016Quantum}. The details about the memory layout are introduced in section \ref{memorylayout}.

Alike the classical program, the QRAM stores the physical variable $\vec{U}^n$ to construct the linear equation at the step n. Besides this, we also prepare the residual vector $\vec{R}$ and a sum tree With the memory layout design, one can construct three quantum subprocedures $O_A$, $O_b$, and $O_l$ required by the QLS. They encode the linear equation as unitary transforms, that is:
\begin{equation}
    O_A|i,k,i',k'\rangle = |i,k,i',k'\rangle|A_{i,k}^{i',k'}\rangle,
\end{equation}
which encodes the Jacobian matrix's element, and
\begin{equation}
    O_b|i,k\rangle = |i,k\rangle|R_{i,k}\rangle,
\end{equation}
which encodes the residual vector's element, and
\begin{equation}
    O_l|i,p\rangle = |i,C_i(p)\rangle,
\end{equation}
which encodes the $p^{\rm th}$ related cell in the difference scheme.

$O_l$ can be implemented directly by the geometry definition input with constant queries. $O_A$ also requires querying the geometry definition to obtain the $O(s)$ number of $U_i$ at the related cells. Implementation of $O_b$ is introduced in section \ref{prepareResidualState}.

With these three quantum subprocedures, QLS outputs a solution $|\vec{u}\rangle=|\frac{\Delta\vec{U}^{n+1}}{\lVert \Delta\vec{U}^{n+1}\rVert_2}\rangle$, a normalized solution of the linear equation.  Now we are able to construct a procedure $P$ which can prepare the $|\vec{u}$ with sublinear time.

Taking $P$ as the input of the $l_\infty$ tomography algorithm, we can obtain a classical vector $\tilde{u}$ which is $\epsilon$-close to the quantum solution $\vec{u}$. This algorithm requires to run $P$ and its controlled version by $O(\frac{\log N}{\epsilon^2})$ many times. As a result, obtaining $\tilde{u}$ requires sublinear time.

Meanwhile, we can apply amplitude estimation\cite{QuantumAmpAmp} to $P$ to obtain the normalization factor $c_l = \lVert \Delta\vec{U}^{n+1}\rVert_2$.  we use $\tilde{u}$ and $c_l$ to update the sum tree and finish this iteration step.

For a steady problem, the computing stops in two cases. One is when the residual is smaller than the convergence limit $\epsilon$, which can be extracted from the top of the tree. Another is when reaching the maximum iteration steps. After stopping, the output of this algorithm is stored in the physical variable area.

\section{Quantum memory Layout for QFVM Algorithm}\label{memorylayout}

To reduce data transfer costs between quantum and classical data, we design a memory layout to efficiently prepare the residual vector state and prepare the oracular input of the quantum linear solver.

\begin{figure*}
    \label{fig:QDS}
    \includegraphics[width=\linewidth]{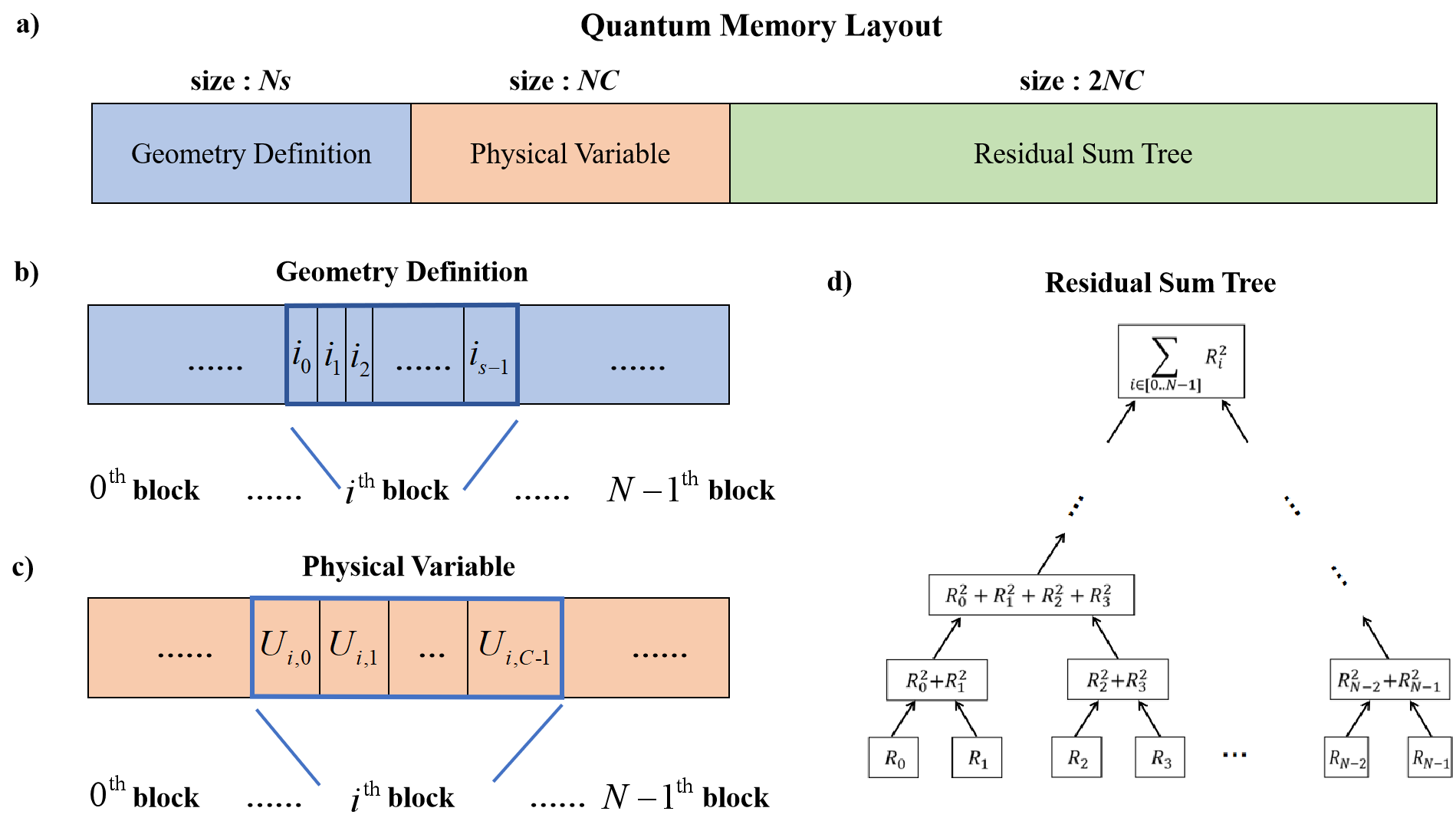}
\caption{Schematic of the quantum memory layout. (a)Three parts of the QRAM memory layout: geometry definition area, which holds the input of the problem; physical variable area holding the $\vec{U}^n$; and the residual sum tree. 
(b)The linear structure of the geometry definition area. This area is formed with N blocks. The $i^{\rm th}$ block holds s related indices where each of $i'=i_k$ ($0\leqslant k<s$) satisfies $C_{i,i'}=1$.
(c)The linear structure of the physical variable area. Each block the physical variables $\vec{U}_{i}$ at the cell i.
(d)The binary tree structure of the residual sum tree. The tree's leaves are the components of the residual vector $\vec{R}^n$. Then for each level, we sum up the square of every two nodes. The tree root is $\lVert \vec{R}\rVert^2$.}
\end{figure*}

The QRAM stores three kinds of data. The first is the geometry data of the problem definition. From this, we can query the connection relation between cells (geometry definition) in quantum parallel. These are constant during the calculation and used for constructing the linear equation. Querying this part is equivalent to this unitary transform:
\begin{equation}
    \mathcal{P}_s|i\rangle|j\rangle=|i\rangle|i_j\rangle.
\end{equation}

The second is the vector of the physical variable at the step n, namely $\vec{U}^n$. From this part we can implement
\begin{equation}
    \mathcal{P}_U|i,k\rangle|0\rangle=|i,k\rangle|U_{i,k}\rangle.
\end{equation}

The third is the residual vector $\vec{R}^n$ and its pre-computed sum. We call it the sum tree. From the bottom to the top of the binary tree, every node stored an integration necessary to prepare the residual state. 
We define each node's address as $a_r(p)$ from the top of the sum tree. $p$ is a binary string where every digit represents the left/right branch with 0/1. For example, $a_r(0)$ is the address of the left child of the root; $a_r(0,1)$ is the right child of the node at $a_r(0)$. Specially, we directly use $a_r$ to represent the root's address. We place all nodes in the QRAM linearly, where each node's address $a_r(p)$ can be computed efficiently. The data contained in address $a_r(p)$ is denoted by $S_R(p)$. Thus we can perform such unitary transform:
\begin{equation}
    \mathcal{P}_R|a_r(p)\rangle|0\rangle=|a_r(p)\rangle|S_R(p)\rangle.
\end{equation}

The second and the third part will be initialized once at the beginning of the calculation. They will be continuously updated through the calculation. The final steady result, which is $\vec{U}$ at the last iteration step, is also stored in them.

The diagram of the memory layout is demonstrated in figure~\ref{fig:QDS}. The functions of this memory layout are as follows:

\begin{itemize}
    \item Initialize the sum tree will classically access the QRAM $O(N)$ times;
    \item With access to the sum tree, one can prepare $|R\rangle$ with $O(\log N)$ times of quantum queries to the QRAM;
    \item If the sparsity of the Jacobian matrix is $s$, updating a single entry of $\vec{U}$ will classical access the QRAM $O(s\log N)$ times to update the sum tree;
    \item The normalization factor between the residual vector and its corresponding quantum state can be obtained with one quantum query;
    \item Evolution from $U^n$ to $U^{n+1}$ will cost $O(C\log^2 N/\epsilon^2)$ time.
\end{itemize}

    

    
    


We clearly state the difference between ``classical access'' and ``quantum query''. Classical accessing means only one data entry is read and modify at a time. Quantum query means a unitary transform is performed, simultaneously extracting many data queries into one quantum register using the superposition addresses.

In the following part, we will show how these features are realized.

\subsection{Initialization}\label{SecInitialization}

The initialization process fills the QRAM as the memory layout scheme shows. The first step is to write in the initial physical variable $\vec{U}$ and fill the tree with the wanted sum.

The initialization is entirely a classical process. Along with writing the data, we should also record the memory layout to quickly obtain the memory address of every data entry in constant time.

There may be some concentrations about whether such $O(N)$ preparation time will cause the vanishment of this algorithm's speedup. We believe after considering the time consumption of initialization, the quantum speedup still preserves. We analyze the three things that contribute to initialization time.

One is the calculation of the residual vector $\vec{R}$. The calculation of this vector exists at every step of the classical FVM in a CFD problem. Even in classical algorithms, this part is not the bottleneck of the time. Our algorithm only calculates the residual once initially, which will consume much less than a classical algorithm does.

The second is the fill of the sum tree. To fill a sum tree only requires repeatedly adding the sum of the square of the residual vector. It is natural to think this process is easier than the calculation of the residual vector.

The third is about the cost of accessing the QRAM classically. As we have mentioned in section \ref{PreliQRAM}, we assume the QRAM has the near capability of RAM, which allows the access to be performed in constant time.


\subsection{Preparation of the residual vector state $|\vec{R}\rangle$}\label{prepareResidualState}

According to the method described in \cite{Grover2002Creating}, the state $|\vec{R}\rangle$ could be prepared efficiently because we have access to all wanted sums of the vector. We pre-compute them in the sum tree, so the preparation can be realized by querying the sum tree $\mathcal{P}_R$.

The first step is to query the tree root and its left child node, then calculate the rotating angle at this step:
\begin{equation}
    |a_r(0)\rangle|S_R(0)\rangle|a_r\rangle|S_R\rangle|\theta\rangle|0\rangle,
\end{equation}
where $S_R(0)=\sum_{i\in [0..N/2-1]} R_i^2$, $S_R=\sum_{i\in [0..N-1]} R_i^2$, $\theta = \arccos\frac{S_R(0)}{S_R}$.

Now perform a conditional rotation and uncompute, we have:
\begin{equation}
    \cos\theta|0\rangle+\sin\theta|1\rangle.
\end{equation}
Add another qubit, perform Hadamard gate on it, we have
\begin{align}
    &(\cos\theta|0\rangle+\sin\theta|1\rangle)\frac{1}{\sqrt{2}}(|0\rangle+|1\rangle) \\
    &=\frac{1}{\sqrt{2}}(\cos\theta|00\rangle+\sin\theta|10\rangle)
\end{align}
Then we iteratively perform the query, computing the rotating angles and conditional rotation. At $k^{\rm th}$ step, we have the state
\begin{equation}
    \sum c_i^k|i\rangle\frac{1}{\sqrt{2}}(|0\rangle+|1\rangle).
\end{equation}

Computing the addresses of $|i,0\rangle$, $|i,1\rangle$ , we obtain the real addresses $a_r(i,0)$ and $a_r(i,1)$ in the QRAM, that is
\begin{equation}
    \sum c_i^k|i\rangle\frac{1}{\sqrt{2}}(|0\rangle|a_r(i,0)\rangle+|1\rangle|a_r(i,1)\rangle,
\end{equation}
then query to the $\mathcal{P}_R$ to obtain the rotating angles and uncompute extra registers. Finally, after performing conditional rotation, we step to
\begin{equation}
    \sum c_i^{k+1}|i\rangle
\end{equation}

Repeatedly performing this process, we can efficiently prepare the residual state $|\vec{R}\rangle = -\sum R_i|i\rangle/\sqrt{\sum_{j\in[0..N-1]}R_j^2}$ with the help of the sum tree.

\subsection{Updating a single entry of the physical variable}

From step n to n+1, the physical variable $\vec{U}$ should be updated. Same as the computing $\vec{R}$ and its sum from $\vec{U}$, only the residual on the related cells would change. From the tree leaves, we change the all residual $R_{i',k'}$ related to the changed $U_{i,k}$ with $C_{i,i'}=1$. After these residual vector entries change, we again compute the sum tree from the leaves to the root and update correspondingly. The number of updated nodes will not exceed the number of the multiplication of the related residual entries $O(s)$ and the number of layers of the sum tree $\log N$.

As a result, the cost of update one entry of $\vec{U}$ is less than $O(s\log N)$.

\subsection{Sampling the solution state and update the QRAM}\label{UpdateQRAM}

The QLS outputs the solution as a quantum state. However, we cannot directly extract it to a classical vector to update $\vec{U}$. We need to cope with two problems: first is to decide the normalization factor of the solution; the second is to convert the quantum state to a classical vector.

The tomography algorithm only produces a normalized vector $\Delta \tilde{U}^{n+1}$. We should also obtain all the normalized factor in the algorithm to get a real update vector of $\vec{U}^{n+1}$. QLS produces two factors. First is $c_b$, which is generated when preparing $|b\rangle$ and can be obtained from the data structure described above. The second is $c_l$ which is generated from the QLS, because matrix inversion is usually not unitary and the raw solution $A^{-1}b$ is not normalized. With amplitude estimation, we can compute the probability $p_l$ and then obtain the factor by $c_l = \alpha \sqrt{p_l}$, where $\alpha$ is a constant in the QLS. Obtaining the normalization factors will not affect the asymptotic time complexity of the algorithm.

Combining these two factors $c=c_bc_l$, we obtain the norm of solution $\lVert\Delta\vec{U}\rVert_2$, which implies the variation updated on the target vector $\vec{U}$ in the CFD solver. When the norm of the variation is smaller than a preset threshold, we can stop the iteration and return the result.

With $l_\infty$ tomography\cite{QuantumDeepCNN} we could efficiently produce an $l_\infty$-close sample $\Delta\tilde{U}^{n+1}$ of a real-valued quantum state $|\Delta\vec{ U}^{n+1}\rangle$ with $O(\frac{\log N}{\epsilon^2})$. This sampling algorithm has a logarithmic dependency on N, enabling each iteration step of our algorithm to run at polylogarithmic time complexity over the input size N.

Updating the QRAM from the sampled vector is also efficient. The $l_\infty$ tomography algorithm produces a sparse classical vector with not more than $O(\log N/\epsilon^2)$ non-zero elements, which means the updating will be performed for less than $O(\log N/\epsilon^2)$ times to update the QRAM $\mathcal{P}_U$ in one iteration step. This results in efficiency in both sampling and updating.


\section{Implementation of Quantum Preconditioner}\label{preconditioner}
The condition number of the linear equation represents to what extent the solution can be affected by the pertubation on the right-hand-side vector. The condition number is defined as:

\begin{equation}
\kappa(\bm{A})=\frac{|\lambda_{max}|}{|\lambda_{min}|},
\end{equation}

where $|\lambda_{max}|$ and $|\lambda_{min}|$ is the maximum/minimum absolute of the eigenvalues of $\bm{A}$. When the condition number is large, we say the equation is ill-conditioned, requiring high precision and time complexity to solve. The time complexity of the classical sparse linear solver has a dependency on the condition number. For example, the time complexity of the conjugate gradient method is $O(\kappa sN\log 1/\epsilon)$. The QLS used in our algorithm also has a linear dependency on the condition number.

Preconditioner is a pre-processing method that can reduce the condition number of the equation. If we have a matrix P such that $\kappa(\bm{PA})<\kappa(\bm{A})$, we can transform this equation as:
\begin{equation}
\bm{A}\vec{x} = \vec{b} \Leftrightarrow \bm{PA}\vec{x}=\bm{P}\vec{b}.
\end{equation}

Preconditioners are constructed from the raw equation, and there have been many types of preconditioners. However, not all classical preconditioners could be directly transplanted to quantum versions. First, the matrix multiplication by the preconditioner should be computed efficiently, namely within $O(\mathrm{polylog} (N))$ time. Second, the preconditioned matrix should also be sparse; otherwise, it cannot be efficiently solved by the QLS. Some preconditioners suitable for QLS have already proposed in \cite{1301.2340} and \cite{1807.04563}. We here display an example preconditioner: the blockwise Jacobi preconditioner, which is widely used in the classical CFD solver. We implement the blockwise Jacobi preconditioner in our algorithm without affecting the asymptotic complexity on the problem size N.

\subparagraph{Apply Jacobi preconditioner to subprocedures}
Jacobi preconditioner uses the inverse of the diagonal block. For the raw linear equation, we construct subprocedures as the input of the QLS. The preconditioned equation has a different matrix and vector; therefore, these subprocedures should be modified.

Let $\tilde{\bm{A}}=\bm{PA}$ and $\vec{R'}=\bm{P}\vec{R}$, where $\bm{P}$ is the Jacobi preconditioner of the matrix $\bm{A}$. The element of the $\bm{P}$ is
\begin{equation}
    \bm{P}_{i,k}^{i',k'}=\delta_{i,i'}B_{i,k}^{i',k'},
\end{equation}
where $\bm{B}_{i,\_}^{i',\_}$ represents the inverse of the block $\bm{A}_{i,\_}^{i',\_}$.

The element of $\tilde{\bm{A}}$ is
\begin{equation}\label{RawElementA_}
    \tilde{A}_{i,k}^{i',k'}=\sum_{j\in[0..N-1];l\in[0..n_{\rm var}-1]}\bm{P}_{j,l}^{i',k'}\bm{A}_{i',k'}^{j,l}.
\end{equation}

The Jacobi preconditioner $\bm{P}$ is blockwise diagonal. We can simplify the equation~(\ref{RawElementA_}) as:
\begin{equation}
    \tilde{A}_{i,k}^{i',k'}=\sum_{l\in[0..n_{\rm var}-1}]\bm{P}_{i,l}^{i,k'}\bm{A}_{i',k'}^{i,l}.
\end{equation}

This implies that computing a single element of $\tilde{\bm{A}}$ requires to queries $n_{\rm var}$ elements of $\bm{A}$. Another fact is that the sparsity matrices of $\tilde{\bm{A}}$ and $\bm{A}$ are the same when they are symmetric to the diagonal line. This is often true because in the difference scheme, $i$ and $i'$ are related so that $A_{i,k}^{i',k'}$ and $A_{i',k}^{i,k'}$ are all non-zero elements.

When it is efficient to implement $O_A$, $O_A'$ will also be efficient to implement. That is
\begin{equation}
O_A'|i,k,i',k'\rangle|0\rangle=|i,k,i',k'\rangle|\tilde{A}_{i,k}^{i',k'}\rangle.
\end{equation}
First we query all elements required for computing the inverse at $(i,i')$ block and compute the inverse, we have
\begin{equation}
    |i\rangle |B_{i,\_}^{i,\_}\rangle_{\rm regs(A)}.
\end{equation}

We use $B_{i,\_}^{i',\_}$ to represent a matrix block with $n_{\rm var}^2$ elements. The subscript ``regs(A)'' mean we require a group of quantum registers to hold this matrix, marked by A.

The corresponding block in $\bm{A}$ is also queried,
\begin{equation}
    |i,i'\rangle |A_{i,\_}^{i',\_}\rangle_{\rm regs(B)},
\end{equation}

Combining two register group A and B, we obtain the wanted element $A_{i,k,i',k'}$. Computing one element requires $n_{\rm var}^2$ times of calls to the $O_A$, namely propotional to $O(s^2)$.

From the above derivation, the sparsity of $\tilde{\bm{A}}$ is same with the $\bm{A}$. Therefore $O_l$ remains unchanged. 

The subprocedure $O_b$ should also be modified. In the original description of the memory layout, the sum tree stores the pre-computed residual vector. In the preconditioned version, the $\vec{R}$ is replaced by $\vec{R'}$. When the $\vec{U}$ changes according to the sampling results, we need to compute the preconditioned residual vector and update the sum tree. Computing any element of $R'_{i,k}$ is still related to all connected cells, which is
\begin{equation}
    R'_{i,k}=\sum_{k'\in[0..n_{\rm var}-1]} \bm{P}_{i,k'}^{i,k}R_{i,k'}
\end{equation}
The complexity of this process is also contributed by computing the diagonal block's inverse of $\bm{A}$.

From the modified sum tree, constructing such $O_b'$ is similar to $O_A'$.

\section{Run Time Analysis}\label{ResourceAnalysis}

The time cost for the QCFD algorithm has two main contributions. One is the cost of the initialization of the memory layout (initialization cost); the other is the time complexity between two iteration steps (evolution cost).

\subparagraph{Initialization cost}
In section \ref{SecInitialization}, we estimate the time cost of the initialization stage. The conclusion is that the initialization stage has $O(N)$ time complexity. However, under the assumption about the QRAM's capability (see section \ref{PreliQRAM}), the initialization cost would not cost much more than the preprocessing stage of the classical FVM. Because the initialization of QCFD only processes once, but classical FVM has to preprocess it as many times as the iteration steps, we believe this cost would not become the bottleneck of the QFVM algorithm.

\subparagraph{Evolution cost}
Every evolution stage. In \cite{Childs2017Quantum}, Childs et al. provide a linear solver algorithm with logarithmic dependence on precision. They show that the query complexity of this algorithm of $O_A$, $O_l$ and $O_b$ are $O\left(s\kappa \mathrm{polylog}(\frac{s\kappa}{\epsilon})\right)$. Now we start to analyze the time complexity of constructing these subprograms from the initial problem settings.

According to the results in the previous sections, the number of queries to QRAM for implementing $O_A$, $O_l$, and $O_b$ is $O(s)$, $O(1)$, and $O(\log N)$, correspondingly. The time complexity of preconditioned $O_A'$ has a multiplier of $O(s^3)$ contributed by computing the inverse of the diagonal blocks of $\bm(A)$. The preconditioned $O_b'$ has the same complexity as $O_b$.

Now consider the time cost of sampling and updating. We run the QLS with $O(\frac{\log N}{\epsilon^2})$ times to obtain an $l_\infty$-close classical vector. This becomes another multiplier to the time complexity of the quantum procedure.

The last multiplier is the cost of querying the QRAM. As we have assumed, the QRAM use $O(\log N)$ time to perform one query. By composing these results, the time complexity of the quantum procedure is
\begin{equation}\label{TimeComplexity}
    O\left(
    \frac{(s^3+\log N)s\kappa\log^3 N}{\epsilon^2}\mathrm{polylog}(s\kappa/\epsilon)
    \right).
\end{equation}

The final step is to update the sum tree. Updating a preconditioned residual tree has two steps. One is to compute the preconditioned residual, where each term will involve in another inversion of the matrix $A$, which is $O(s^3)$; the other is to update the tree from bottom to the top, this involves $O(\log N)$ times for one change in the bottom of the tree. While at most $O(\frac{\log N}{\epsilon^2})$ terms of $\vec{U}$ changes, the time cost of updating the tree is $O(s^3 \log^2 N/\epsilon^2)$. The total time complexity is the addition of the quantum and the classical procedure. Because the quantum procedure's complexity is asymptotically greater than the classical's, we conclude that the evolution time cost has the time complexity shown in equation~(\ref{TimeComplexity}).

The classical counterpart's time complexity is $O(Ns\kappa \log 1/\epsilon)$ when using CG as the linear solver. Our algorithm outperforms the classical algorithm on the problem size's dependency but has worse performance when the problem requires high precision. When the problem size $N$ and the requirement of the precision $\epsilon$ has such relation $N\gg 1/\epsilon^2$, the quantum algorithm will potentially have better performance on time.

\section{Numerical Simulation}

To prove our algorithm's effectiveness, we performed simulation on a test case to find whether a big size problem with low error sensitivity exists. 

An open-source classical CFD simulation software, SU2\cite{economon2016su2}, is selected as the example. We appended quantum error in the simulation process, output the evolution history, and compared it to the classical solver. The error is implemented by biasing the solution vector with a randomized error to emulate the sampling process of the $l_\infty$ tomography. Except for the error, we preserved all physical problem configurations, including the mesh setting and physical parameters like temperature or Mach number. The multigrid option was turned off so that the linear equation will only be solved once in every iteration step. The linear solver parameters are chosen to be as precise as possible to emulate the case that the quantum linear solver produces the result accurately.




We select the inviscid flow around Onera M6 airfoil as the test case, which has 108396 grid points, and the physical governing equation is the three-dimensional Euler equation. We compare the classical result with different error settings: from 5e-2 to 1e-5. The results are displayed in figure 2. When the error is set to 5e-2 quickly diverges. Except that, all cases converge correctly. The black line is the classical baseline. The maximum convergence error $\epsilon_1$ of this case is between 5e-2 and 3e-2, where quantum advantage preserves.

To simulate the quantum effect, we biased the solution vector with a specific error to emulate the sampling process of the $l_\infty$ tomography. This is implemented as a postprocessor of the linear solver. 

\begin{figure*}
    \label{fig:Onera}
    \includegraphics[width=\textwidth]{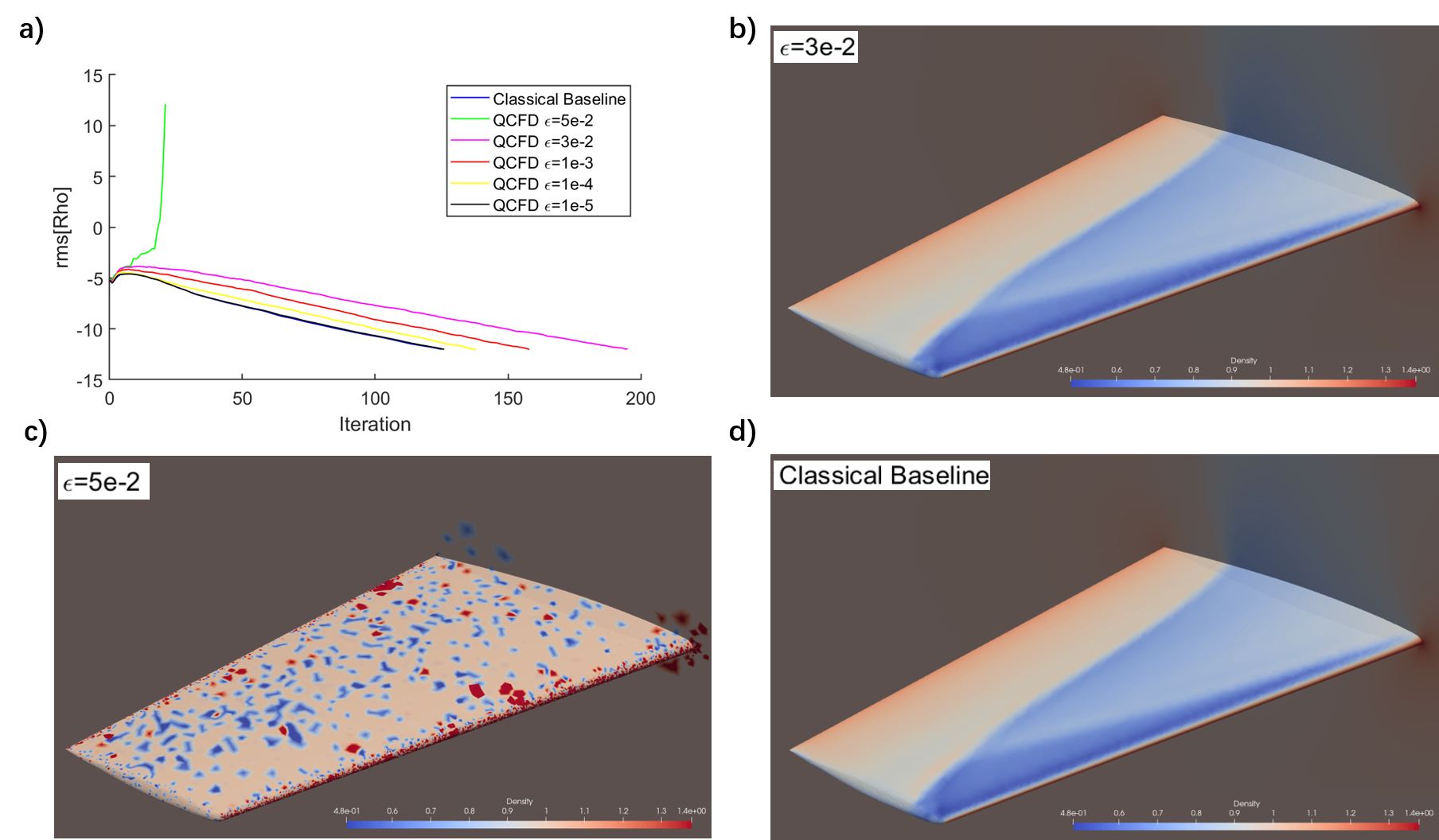}
\caption{Results of the simulation. The test case is the three-dimensional inviscid flow around the Onera M6 airfoil. a) Convergence history of the test case of the quantum solver with different error setting, compared to the classical solver as the baseline (use BCGStab as the linear solver). The error is set from 5e-2 to 1e-5. Except $\epsilon$=5e-2, all other cases converge correctly. The maximum convergence error is approximately 3e-2. b) Flow density around the airfoil at quantum error $\epsilon$=3e-2. This result is the same as the classical solver's result displayed in subfigure d). c) Bad solution at $\epsilon$=3e-2. The flow density at 20 iteration steps is not correctly computed. d) The solution output by the classical solver.}
\end{figure*}

We focus on the convergence history of test cases and define two kinds of error thresholds. The first is ``maximum convergence error $\epsilon_1$'', the maximal value which allows the convergence. The problem will quickly diverge if the error is larger than the first threshold. The second is ``maximum stable error $\epsilon_2$'', where the problem will have the same convergence history if the error is less than this threshold. The convergence speed will be gradually slow when the error gets larger between the first and the second threshold. If the maximum stable error is large, we define this kind of problem as ``quantum friendly'' because the quantum algorithm is much likely to run faster than the classical. In this case, the problem size is large, satisfying $N\gg 1/\epsilon^2$. This provides evidence that quantum advantage can be realized in CFD problems.


\section{Error Analysis}
The time complexity of the QFVM has better performance on the number of grid cells $N$ but worse on the precision $\epsilon$, which implies that the problem size should be large enough to show the quantum advantage. On the other side, the numerical simulation shows that the precision should be small. Otherwise, the time integration will not converge. There is the problem: if the precision requirement has some dependency on the problem size, the quantum acceleration will decrease or even vanish. In this section, we will provide evidence that the precision will not grow with the problem size.

First, we calculate the total error generated by the quantum sampling with error bound $\epsilon$ specified. At one step, we define the physical variable $\vec{U}$ and its update $\Delta \vec{U}$. In QFVM, the quantum process outputs a quantum state $|u\rangle$ which is propotional to $\Delta \vec{U}$,
\begin{equation}\label{eq:defu}
    \Delta \vec{U} = \lVert \Delta \vec{U}\rVert_2 u.
\end{equation}

The $l_\infty$ tomography outputs a classical vector $\tilde{u}$ which is $\epsilon$-close to $u$. At any index $i$, we have
\begin{equation}
    u_i = \tilde{u}_i + e_i,
\end{equation}
where the error term $|e_i|<\epsilon$.

Now we consider the amplitude of the $e_i$. When performing $l_\infty$ tomography, the output vector is a sample from the multinomial distribution where the sampling number $M = C\log N/\epsilon^2$ and the probability distribution ($|u_0|^2$, $|u_1|^2$, ... $|u_{N-1}|^2$). At any term, the standard error of such sample is:
$\sigma_i = \sqrt{M|u_i|^2(1-|u_i|^2)}$. When $|u_i|$ is small enough, we have $\sigma_i\sim |u_i|\sqrt{N}$. Now we assume the error $e_i$ is approximatedly linear dependent on the standard error $\sigma_i$, thus we have
\begin{equation}\label{eq:e}
    e_i = O(\sigma_i) \sim O(|u_i|\sqrt{N}).
\end{equation}

The update vector output by the QFVM should be multiplied by $\lVert \Delta \vec{U}\rVert_2$. As a result, the total error will be amplified by this coefficient.
\begin{equation}\label{eq:E}
    E_i = \lVert \Delta U\rVert_2 e_i.
\end{equation}

Compare two cases describing the same problem where one has $N$ cells and the other has $kN$ (mark the variables with extra prime, e.g. $u'$). We can assume the distribution of $\Delta \vec{U}$ and $\Delta\vec{U}'$ is the same because the physical characteristic does not change. From this, we have
\begin{equation}
    \lVert \Delta U'\rVert_2^2 = k\lVert \Delta U\rVert_2^2,
\end{equation}
because only the vector size changes to $k$ times. From the definition of $u$ (equation~(\ref{eq:defu})), this results in the decrease of the amplitude of the $u$, i.e.
\begin{equation}\label{eq:ui}
    u_i' = \frac{1}{\sqrt{k}}u_i.
\end{equation}

Combining equation~(\ref{eq:e}), (\ref{eq:E}) and (\ref{eq:ui}), we obtain that $E_i = O(U_i)$. This result implies that the total error generated by the quantum sampling will not change over the problem size $N$.

\section{Conclusion}
This paper developed a quantum version of the finite volume method (QFVM) for solving the CFD problem with substantial speedup. Identical to the classical FVM, the QFVM is iterative and can output the evolution history of the fluid flow in the computing space. The input and output of the algorithm are both classical data. In conclusion, we give the time complexity of between each step is $O\left(\frac{(s^3+\log N)s\kappa\log^3 N}{\epsilon^2}\mathrm{polylog}(s\kappa/\epsilon)\right)$, where $N$ is the number of cells (the problem size), $\kappa$ is the condition number of the linear equation, $s$ is the sparsity of the Jacobian matrix, and $\epsilon$ is the precision threshold of the output. Compared to the classical solver's best time complexity, which has a linear dependency on the grid size $N$, this algorithm is exponentially faster. The speedup would be significant when we choose an extremely large $N$,  which allows the quantum computer to solve complex CFD problems, such as tackling a larger space or finer mesh. 

We use the QLS to accelerate the solution of the linear equation. However, previous works often ignore the cost of transferring data between quantum and classical computers. To achieve an efficient transfer, we design a memory layout based on the QRAM. The memory layout stores the problem configuration and the internal result. With a quantum parallel query, one can implement quantum subprocedures necessary for the QLS; it is also updated efficiently when the algorithm iterates over steps.

Numerical simulations are conducted to check whether the final result is affected by the quantum error. The problem size of the test case is $1e5$, and the QCFD solver converges correctly at $\epsilon$=3e-2, after around 200 steps. This result shows that CFD could have strong error tolerance with large problem size; therefore, the quantum advantage preserves.

Our future work will focus on how precision affects the final result and the convergence history, and how to optimize it. We believe that the quantum computer will show its advantage in solving a more complex CFD problem shortly.

\begin{acknowledgments}
This work was supported by the National Natural Science Foundation of China (Grants Nos. 11625419), the National Key Research and Development Program of China (Grant No. 2016YFA0301700),  the Strategic Priority Research Program of the Chinese Academy of Sciences (Grant No. XDB24030600), and the Anhui Initiative in Quantum Information Technologies (Grants No. AHY0800000).
\end{acknowledgments}

\bibliography{apssamp}

\end{document}